# Selection rules of twistronic angles in 2D material flakes via dislocation theory

Shuze Zhu[1, 2], Emil Annevelink[2], Pascal Pochet[3], Harley T. Johnson[2,4]

(1) Department of Engineering Mechanics, Zhejiang University, Hangzhou, 310027, China
(2) Department of Mechanical Science and Engineering, University of Illinois at Urbana-Champaign, Urbana, IL 61801, USA
(3) Department of Physics, IriG, Univ. Grenoble Alpes and CEA, F-38000, Grenoble, France
(4) Department of Materials Science and Engineering, University of Illinois at Urbana-Champaign, Urbana IL 61801 USA

**Abstract:** Interlayer rotation angle couples strongly to the electronic states of twisted van der Waals layers. However, not every angle is energetically favorable. Recent experiments on rotation-tunable electronics reveal the existence of a discrete set of angles at which the rotation-tunable electronics assume the most stable configurations. Nevertheless, a quantitative map for locating these intrinsically preferred twist angles in twisted bilayer system has not been available, posing challenges for the on-demand design of twisted electronics that are intrinsically stable at desired twist angles. Here we reveal a simple mapping between intrinsically preferred twist angles and geometry of the twisted bilayer system, in the form of geometric scaling laws for a wide range of intrinsically preferred twist angles as a function of only geometric parameters of the rotating flake on a supporting layer. We reveal these scaling laws for triangular and hexagonal flakes since they frequently appear in chemical vapor deposition growth. We also present a general method for handling arbitrary flake geometry. Such dimensionless scaling laws possess universality for all kinds of two-dimensional material bilayer systems, providing abundant opportunities for the on-demand design of intrinsic "twistronics". For example, the set of increasing magic-sizes that intrinsically prefers zero-approaching sequence of multiple magic-angles in bilayer graphene system can be revealed.

**Keywords**: 2D materials, vdW interfaces, Energy landscape, Twistronics, Dislocations

**INTRODUCTION**

Tuning electronic states in twisted two-dimensional (2D) van der Waals (vdW) layers represents an emerging field known as "twistronics" [1], where controlling the interlayer rotation angle is critical [2-4]. Examples include the correlated insulator states and superconductivity in magic-angle (i.e., angle with flat band formation) twisted bilayer graphene [5-10], angle-dependent conductivity in twisted bilayer graphene interface [11], as well as tunable photoluminescence spectra [12] and band gaps of bilayer $MoS_2$ over twist angles up to tens of degrees [13]. Existing experimental control on interlayer rotation angle usually begins with transferring a top layer flake onto another supporting layer with relative twist imposed [12,14-17].

Nevertheless, the twist techniques which rely on imposing well-controlled global rotation extrinsically [5,7,9,18] suffer from the fact that not every angle is intrinsically favorable. Recent experiments on rotation-tunable electronics [1], where atomic force microscope tip is used to rotate the top flake over a range of angles [1,19-21], confirm the existence of a discrete set of twist angles at which the top flake is more stable and resistant to rotation. Energetically, these angles are associated with local minima of rotating interface energy landscape [1,22]. In addition, if the applied global twist angle is not intrinsically favorable, local energetic relaxation over twisted interface may result in disorder [6] of local twist angle, the mapping of which may be difficult. The mapping of local twist angle disorder in magic-angle graphene has only recently been demonstrated using a nanoscale on-tip scanning superconducting quantum interference device [6]. Therefore, on-demand design and investigation of twisted structures that intrinsically favor definite desired twist angles still remain challenging. A particular example is the difficulty in accessing the higher order zero-approaching sequence of multiple magic-angles of graphene [8] (e.g, 0.5°, 0.35°, 0.24°, and 0.2°), where interesting physics might arise.

Ideally, one could design a system in which the desired twist angles coincide with angles at which the interface energy is at a local minimum. It would then be possible to envision an intrinsic twisted structure, where the top layer flakes intrinsically favor the desired twist angles. For example, one could envision a top layer flake in a twisted bilayer graphene system (e.g., fabricated from either chemical vapor deposition growth or etching if the right geometry of flake is known) intrinsically prefers to be in a magic-angle rotation relative to the bottom layer. The robustness and precision of twist angle are protected by intrinsic interface energetics. Then any perturbation (e.g. tip manipulation) would easily deliver the system to that stable state of desired twist angle with certainty. However, missing in this picture is a quantitative map for locating angles of local energy minima for various twisted 2D material system, which is critical to address the challenge of on-demand design of intrinsic "twistronics".

Here we reveal this map in the form of geometric scaling laws for a wide range of intrinsically preferred twist angles as a function of only geometric parameters, such as size and shape, for a rotating flake. Our analysis starts with the generic geometric description of the interface in a twisted bilayer system, and then provides a description of the associated geometrically necessary dislocation networks, from which the geometric scaling laws gradually emerge. Using triangular and hexagonal shapes as examples (since they are most frequently seen in chemical vapor deposition growth), the as-found scaling laws are completely dimensionless and geometric, and thus possess universality for all kinds of twisted 2D vdW bilayers that are governed by the mathematically-precise geometry of moiré periodicity. In light of such scaling laws, the design and investigation of intrinsic "twistronics", which intrinsically favor desired twist angles, are feasible and possible. Given the strong connection of twist angles to various emerging condensed matter properties in 2D material layers, these scaling laws offer rich opportunities.

**RESULTS AND DISCUSSION**

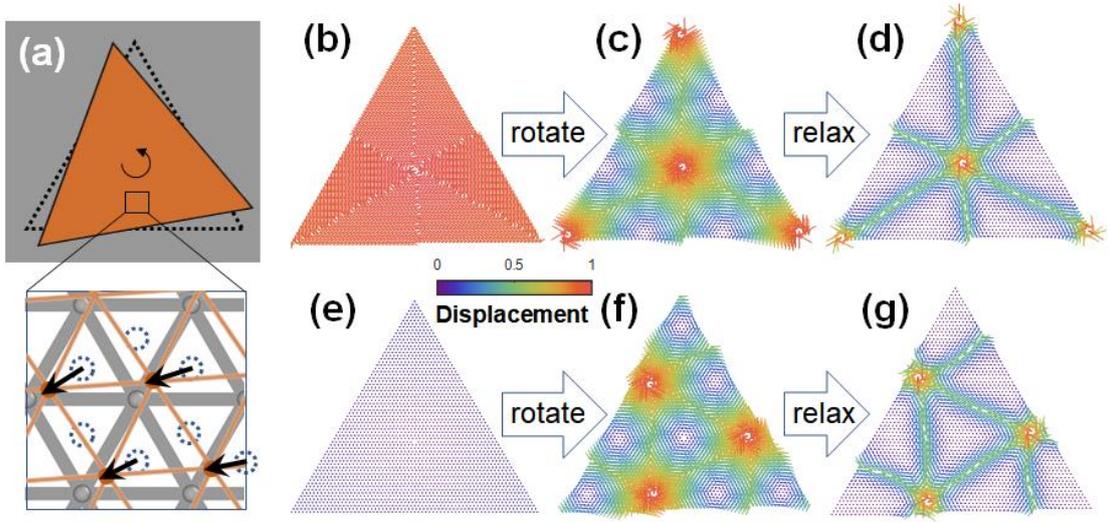

FIG. 1. (a) Geometric characterization of generic displacement vectors during rotation, defined as in-plane deviation of the atoms in the top layer (flake) from their nearest equilibrium sites (dashed circles) in the bottom layer. (b) For initial AA stacking, every atom has the largest possible displacement. (c) As it rotates, periodic clusters of large displacements emerge. The paths connecting these large-displacement clusters demarcate the flake into low-displacement regions. (d) Generic relaxed configuration. (e-g) is for initial AB stacking, where initially every atom has the smallest possible displacement. The displacements are normalized by the largest possible magnitude, and are colored according to their magnitude.

A twisted bilayer 2D material interface can be understood as assemblies of periodic regions of interlayer commensurability and incommensurability. These assemblies, known as superlattices or moiré patterns, result from a purely geometric effect of rotation. For any amount of rotation, the moiré wavelength is encoded in the following geometric function [23,24] of the material lattice constant $a$, and rotation angle $\theta$:

$$\lambda(\theta) = \frac{a}{\sqrt{2 - 2cos\theta}} \quad (1)$$

A displacement vector can describe the direction and magnitude of the deviation from nearest equilibrium location for every atom in either layer, as illustrated using a generic bilayer system of triangular lattices [Fig.1(a)]. This geometry could represent, for example, the triangular sublattice of bilayer graphene system. Or, it could represent the triangular lattice of interfacial sulphur layers of bilayer $MoS_2$ system. Without loss of generality in the analysis that follows, the top layer is assumed to be triangular in shape with equal side length, while the bottom layer is assumed to be much larger. The rotation axis of the top flake passes through its symmetrical center. Then for each atom in the top flake, the displacement vector is defined as the in-plane deviation from the nearest minimum energy site on the bottom flake [Fig. 1(a)].

Fig. 1(b-c) shows the evolution of displacement vectors, colored according to their magnitudes, for a representative rotation from the initial AA stacked configuration. Before rotation, all the atoms on the top flake have the same large-displacements due to the nature of AA stacking. After a typical rotation, the average displacement magnitude (i.e., the central range of the contour scale) is distributed along paths whose intersections are occupied by clusters of atoms with large displacement. These large-displacement junctions in Fig. 1(c) appear at the flake center as well as near the flake edge. The regions separated by these paths are filled with small-displacement atoms. The displacement vectors, defined as above, do not contain any in-plane relaxation, and can therefore represent the weak interfacial coupling case. For strong interfacial coupling, the displacement vectors are expected to further evolve. Since relaxation would try to pull each atom towards its nearest minimum energy site, by utilizing a generic interfacial potential (see Section 1 in supporting information), the generic relaxed configuration of displacement vectors can be simulated [Fig. 1(d)]. Together with Fig. 1(c), it can be observed that regardless of the interfacial coupling strength, the large-displacement regions are distributed along the paths whose intersections are the large-displacement junctions. This observation also holds true for the rotation from initial AB stacked configuration [Fig. 1(e-g)]. It is clear that the creation of these paths, whose intersections are the large-displacement junctions, is a purely geometric effect of rotation. We next use bilayer systems of graphene and $MoS_2$ as examples to show that such geometric effects are topological and represent geometrically necessary dislocations.

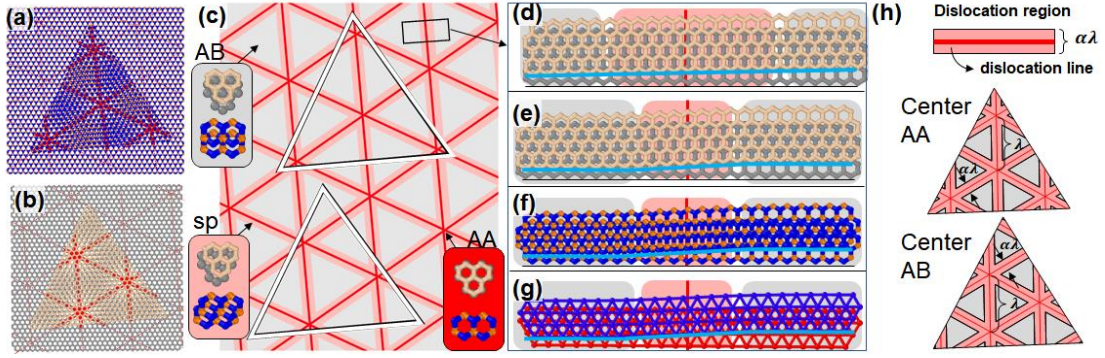

FIG. 2. MoS$_2$ (a) and graphene (b) finite flakes rotated by 5° produce an atomic convolution image. The dislocation network (red lines on the background) is superimposed onto the atomic moiré pattern. (c) Twisted Moiré pattern highlighting the periodicity of different stacking shown in insets (AB, sp, and AA in grey, pink, and red respectively). White solid triangles outline flake region as in (a-b). (d-g) Schematics showing transition of AB-sp-BA stacking across the dislocation line to highlight its topology. Light blue thick lines are visual guides to show the Burgers circuits and relaxations. Dislocation topology of graphene for a pure twist (d) and relaxed (e) configuration. Dislocation topology of MoS$_2$ for all atoms (f) and just interfacial sulfur atoms (g). (h) Dislocation line model for AA and AB center stackings. The dislocation line region carries a finite width $\alpha\lambda$, where $\alpha$ is a constant and $\lambda$ is the moiré wavelength.

Fig. 2 illustrates how the topology of the displacement vectors characterizes the geometrically necessary partial dislocations in bilayer systems of graphene and MoS$_2$. Fig. 2(a-b) shows atomic convolutions of MoS$_2$ and graphene homo-bilayers rotated by 5°, for the as-shown flake size. The rotation introduces a triangular moiré pattern highlighted by red lines on the background. Fig. 2(c) further abstracts the atomic convolutions of the moiré topology into grey regions, representing the ground-state (termed as AB stacking), separated by pink regions, representing the saddle point [25] (termed as sp stacking). The junctions of pink regions assume AA high-energy stacking, where atoms from each layer are on top of each other. Such moiré topology is purely geometric. For our analysis, we center the flakes on initial AA/AB stacking to show the broad applicability of the dislocation interpretation.

The topological character of a boundary line separating stacking regions is defined by how the atomic structure changes across it. Fig. 2 (d-e) shows the left-to-right transition from AB to sp to BA for a perfect twist (no interface relaxation) and a shear boundary (with interface relaxation) in bilayer graphene. The interface relaxation changes displacements but maintains the topology (AB-sp-BA) of the boundary line. The topological character is defined using a Burgers circuit (light blue lines are visual guides, also see Section 2 in supporting information), which shows that the Burgers vector is parallel to the boundary line separating stacking regions [26]. Therefore, the displacement topology (e.g., Fig. 1) represents partial screw dislocations with Burgers vectors parallel to their line direction. They are partial dislocations because they separate equivalent (AB/BA) not identical (AB/AB) stacking regions. For $MoS_2$, the topological analysis is shown for all atoms [Fig. 2(f)] and for just interfacial sulfur atoms [Fig. 2(g)]. In either case, the Burgers circuit again defines the displacement topology as partial screw dislocations. Therefore, both twisted bilayer system of graphene and $MoS_2$ can be equivalently described as networks of dislocations [27,28].

The above dislocation topology can be generalized to a generic dislocation line model as described in Fig. 2(h), where the dislocation line area stripe is colored pink, the central dislocation line is in red, and the grey region is assumed dislocation free. The distance between the dislocation junction is $\lambda$ as in equation (1). The dislocation line region carries a finite width $\alpha\lambda$, where $\alpha$ is assumed a constant. Although decreasing $\alpha$ represents increasing interfacial interaction, in which case the commensurable region (i.e., the small-displacement region) tends to expand more in area and thus reduce the width of dislocation line region, $\alpha$ is shown later to be independent of the angles of local energy minima. Because the location of dislocation junction during rotation can be pre-determined mathematically, one can thus track the change in the total area of the geometric union of these dislocation line regions within the flake boundary, which will be calculated numerically, as a function of rotation angle. Fig. 3(a) shows a sequence of dislocation region configurations during rotation with initial AA stacking. The flake center is a junction due to the nature of AA stacking [Fig. 1(c)]. Before rotation, the entire flake is in a dislocation region due to the very large junction spacing, consistent with Fig. 1(b). As rotation increases, the junction spacing $\lambda$ decreases. The rotation thus geometrically drives dislocation junctions, which are previously exterior to the flake, across the boundary to reside inside the flake.

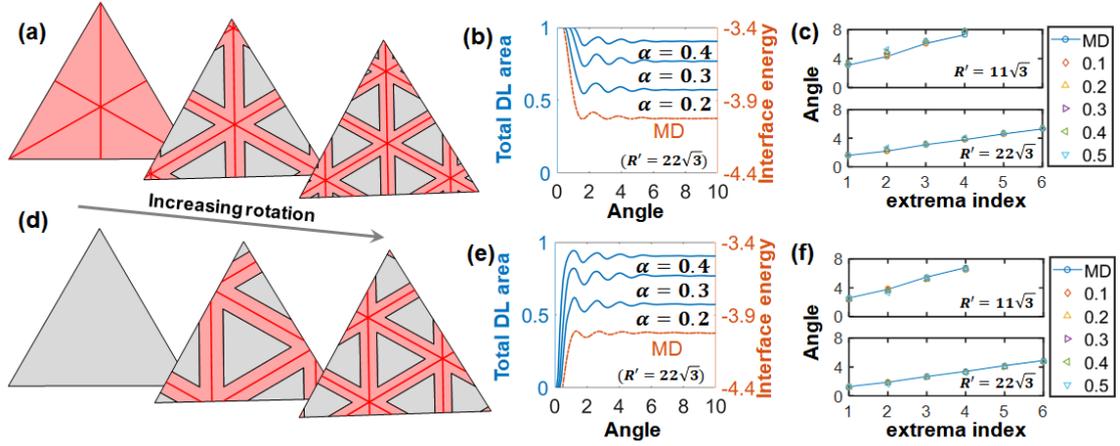

FIG. 3. (a) Dislocation line region configurations during the rotation of a flake with initial AA stacking. (b) For initial AA stacking, the comparison of the normalized total dislocation line area of varying line width (top three blue solid curves, $\alpha$=0.4,0.3,0.2) with the normalized interface energy from MD simulation (bottom brown dashed curves) for a normalized flake size. The angles of local extrema are in good agreement. (c) For initial AA stacking, the comparison in angles of local extrema of the total dislocation line area with the interface energy from MD simulations, for two normalized flake sizes and five sets of $\alpha$, which show good agreements. (d-f) is the parallel version of (a-c) but for initial AB stacking.

The evolution of the total area of the geometric union of these dislocation line regions exhibits a sequence of extrema during rotation [Fig. 3(b)]. More importantly, these angles of area extrema, are insensitive to dislocation line width [Fig. 3(b-c)], which is generically coupled to various material-specific interlayer relaxation. Furthermore, these angles of area extrema correspond to angles of interface energy local extrema [Fig.3(b)]. While the angles of local extrema vary for different flake sizes, the insensitivity to the dislocation line width remains [Fig. 3(c)]. These observations, suggesting the nature of insensitivity for the angles of local extrema to various degrees of relaxation, are clear from the comparison [Fig. 3(b)] of the total dislocation line area with the interface energy calculated by MD simulation [29,30] for a rotating flake with triangular lattices (see Section 3 in supporting information), and from the comparison [Fig. 3(c)] among two flake sizes and five sets of dislocation line width. All these comparisons are made in terms of normalized quantities defined as follows. The total dislocation line area is normalized by the area of the flake. The interface energy is normalized by the interatomic interaction strength in the assumed Lennard-Jones vdW potential (i.e., normalized by the $\epsilon$ in $V(r) = 4\epsilon(\frac{\sigma^{12}}{r^{12}} - \frac{\sigma^6}{r^6})$). The normalized flake size $R' = R/a$ is defined as the vertex-to-center distance of the flake $R$ divided by the implicit material lattice constant

$a$ that appears in equation (1). Hereinafter we will employ above normalizations and notations unless noted otherwise. Fig. 3(d-f) further shows the same results as Fig. 3(a-c) but for the initial AB stacking case. Because MD simulations are designed to enable the sampling of interface energy at fixed imposed angle [31,32], which considers no in-plane relaxation, the convergence of angles of local extrema (e.g., Fig.3(c, f)) to those calculated from MD simulation thus suggests that the insensitivity applies to the weak coupling limit. Therefore, the insensitivity is to be maintained over a spectrum of interlayer couplings in various material systems.

The nature of insensitivity in the angles of local extrema to various degrees of relaxation hints at the existence of a geometric invariant quantity. In other words, these angles should be able to reveal themselves from a purely geometric calculation (i.e., free from any material-specific relaxation). To see this, one suitable calculation can be done in the framework of purely plane-geometric unrelaxed displacement [Fig. 1(a)]. We seek to calculate the mean displacement, defined as the average of the magnitude of displacements among all the atoms on the top flake during rotation. Fig. 4(a) shows the agreement of angles of local extrema in the evolution of mean displacement with the normalized total dislocation line area during rotation for the same normalized flake size as in Fig. 3(b, e), and for both initial stackings. These agreements confirm that the angles of extrema for total dislocation line area (regardless of the dislocation line width), the interface energy, and the mean displacement converge. Therefore, the geometric invariant quantity that governs the angles of local extrema can be associated with displacement field [Fig.1(a)].

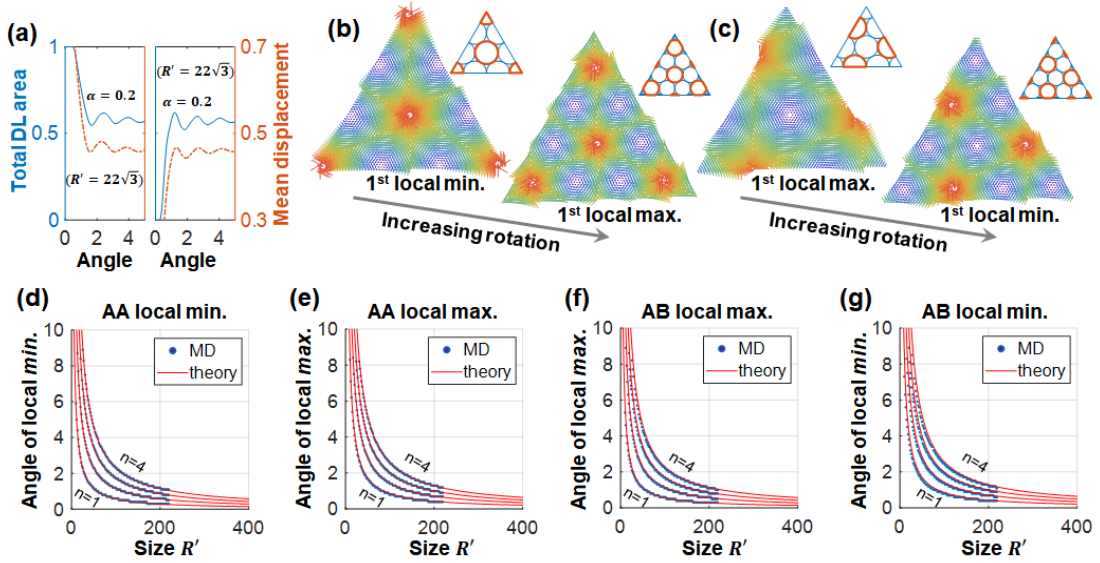

FiG. 4. (a) The correlation of angles of local extrema for normalized total dislocation line area (from Fig. 3) with the mean displacement during rotation. Left panel: AA stacking; right panel: AB stacking. (b-c) The displacement maps correspond to 1st local extrema in mean displacement for rotation from initial AA stacking (b) or AB stacking (c). The flake can be divided into two sub-regions with large and small mean displacement, bounded by red and blue circles as shown in the top-right sub-image. (d-g) The comparison of the angles of $n^{th}$ local energy extrema (obtained from MD simulation of twisted bilayer graphene) with the prediction from geometric scaling theories (equation (4-5), $n=1,2,3,4$), as a function of normalized flake size, for both types of center stackings.

We then work with geometry of displacement field to reveal the geometric scaling laws of angles of local extrema. Taking advantage of the geometric insensitivity, we assume the geometry of large-displacement regions and low-displacement regions to be circles of equal diameter that are tangential to each other, as shown in the inset in Fig. 4(b). We find that this assumption significantly simplifies the derivation of a first order geometric approximation to the scaling laws for the angles of extrema. Quantitative calculations (see Section 4 in supporting information) confirm that, for a triangular flake with initial AA stacking, when the centers of the red circles are on the flake boundaries, the ratio of the area occupied by the blue circles (i.e. low-displacement regions) to the area occupied by the red circles (i.e. large-displacement regions) reaches its maximum, thus corresponding to minimum mean displacement. This geometric condition can be approximated by $\lambda = R = aR'$. Considering equation (1), one obtains the scaling law for the first local minimum angle for rotation of a triangular flake from initial AA stacking:

$$\theta = \arccos(1 - \frac{1}{2R'^2}) \qquad (2)$$

Similarly, when major fractions of the red circles are enclosed within the flake boundary, the area ratio of blue circles to red circles reaches its minimum, thus corresponding to the maximum mean displacement. This geometric condition can be approximated by $\left(1 + \frac{1}{2}\right)\lambda = R = aR'$, which leads to the scaling law for the first local maximum angle for rotation of a triangular flake from initial AA stacking.

$$\theta = \arccos(1 - \frac{9}{8R'^2}) \qquad (3)$$

However, the nature of extrema is reversed for the initial AB stacking case [Fig. 4(c)]. This is caused by the difference in the geometric dislocation topology (i.e., geometric distribution pattern of blue and red circles). Nevertheless, the mathematical form of the geometric condition remains unchanged, which can be easily confirmed by inspecting the geometry.

Equations (2-3) can be easily generalized for a sequence of angles of local extrema by considering periodicity of dislocation junctions (see Section 4 in supporting information). Namely, the $2n - 1$ th local extreme energy state ($n^{th}$ min. for AA; $n^{th}$ max for AB) satisfies $n\lambda = R = aR'$, which leads to angles of local extrema as

$$\theta = \arccos(1 - \frac{n^2}{2R'^2}) \qquad (4)$$

The $2n$ th local extreme energy state ($n^{th}$ max. for AA; $n^{th}$ min for AB) satisfies $\left(n + \frac{1}{2}\right)\lambda = R = aR'$, which leads to angles of local extrema as

$$\theta = \arccos(1 - \frac{(2n+1)^2}{8R'^2}) \qquad (5)$$

Equations (4-5), applicable to triangular flakes, are summarized in Table 1. The explicit ordering of extrema can be easily tracked by considering the initial stacking. For example, for initial AA stacking, the first extreme has to be local minimum because initially the mean displacement is the maximum [Fig. 1(b)]. Also note that higher-order angles of local extrema are associated with more spatial periods of moiré pattern within the flake (see Section 4 in supporting information).

The accuracy of these scaling laws (equations (4-5)) can be tested against the large-scale MD simulation for twist bilayer graphene (see Section 3 in supporting information), where the sampling on angles of local extrema is performed by calculating the interface energy at fixed imposed angle for a wide range of sizes. The comparisons are displayed in Fig. 4(d-g), where the MD simulation data have been converted to normalized form. The strong agreement between the scaling laws (equations (4-5)) with a specific material example (i.e., twisted bilayer graphene), over a wide range of sizes, and up to several orders of extrema, further attests to the universality of these geometric scaling laws and our analytical approaches.

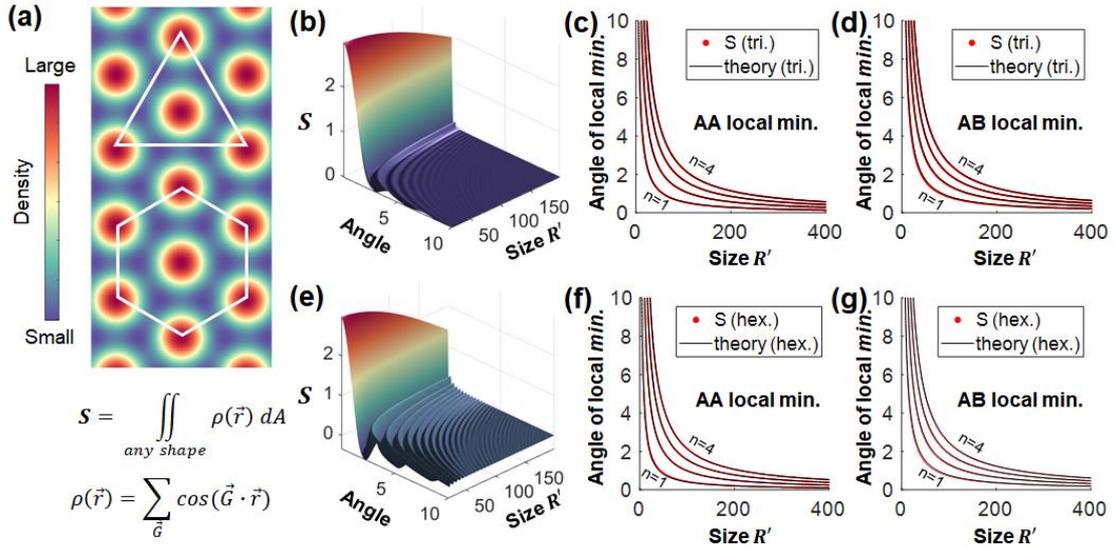

FIG. 5. (a) Generic periodic density function respecting the spatial periodicity of the dislocation junctions. Then the total quantity that is associated with the summation of the general density function over a region can be expressed as integral (termed as $S$ integral), where the integration region can be an arbitrary flake shape. (b) The surface plot of $S$ as a function of rotation angle and normalized flake size, for triangle flakes with initial AA stacking. (c-d) The comparison of the predicted angles of local minimum by $S$ integral and scaling theories (equations (4-5)), for AA and AB initial stacking, respectively. (e) The surface plot of $S$ as a function of rotation angle and normalized flake size, for hexagon flakes with initial AA stacking. (f-g) The comparison of the predicted angles of $n^{th}$ local minimum by $S$ integral and scaling theories (equations (6-7), $n=1,2,3,4$), for AA and AB initial stacking, respectively.

Finally, we consider arbitrary flake shapes. Since the normalized dislocation junction spacing $\lambda/a$ only depends on rotation angle (see equation (1)), it is thus also a fundamental geometric invariant quantity. Therefore, we must also be able to directly extract the angles of local extrema from it. To see this, we resort to Fourier series expansion in the following procedure, which enables the analysis of arbitrary flake shapes [Fig. 5(a)]. First a generic periodic density function is expressed as $\rho(\vec{r}) = \sum_{\vec{G}} \cos(\vec{G} \cdot \vec{r})$, where $\vec{G}$ denotes reciprocal lattice vectors coupled to the spatial periodicity of the dislocation junctions. Then the total quantity that is associated with the summation of the generic density function over a region can be expressed as $S = \int \rho(\vec{r}) \, dA$, where the integration region can be an arbitrary flake shape. Here we call this integral as summation integral, or $S$ integral. Then by inspecting the value of $S$ integral in the space of shape size and rotation angle, one can easily identify the local angles of extrema as a function of size, and thus delineate the scaling law. We next demonstrate the application of $S$ integral to triangular and hexagonal flakes (see Section 5 in supporting information).

Fig. 5(b) shows the surface plot of $S$ integral for a triangle of initial AA stacking as a function of rotation angle and normalized flake size, where a sequence of trenches on the surface is identified. Extracting the angles and sizes corresponding to the trenches (i.e., local minimum), and plotting them against the scaling theory (i.e., equation (4)), one can see again the good agreement [Fig. 5(c)]. Similarly, the $S$ integral for a triangle of initial AB stacking produces agreeing angles of local minimum [Fig. 5(d)] with scaling theory (i.e., equation (5)). Fig. 5(e) shows the surface plot of $S$ integral for a hexagon of initial AA stacking, where more salient features of trenches emerge in terms of depth as compared with that of triangle case [Fig. 5(b)].

To enable direct comparison, the scaling laws for hexagon shape are revealed using same techniques as before (see Section 4 in supporting information). For a hexagonal flake, the $2n - 1$ th local extreme energy state ($n^{th}$ min. for AA; $n^{th}$ max. for AB) satisfies $\left(n - \frac{\sqrt{3}}{6}\right) \lambda = R = aR'$, which leads to angles of local extrema as

$$\theta = \arccos(1 - \frac{(2\sqrt{3}n - 1)^2}{24R'^2}) \qquad (6)$$

The $2n$ th local extreme energy state ($n^{th}$ max. for AA; $n^{th}$ min. for AB) satisfies $\left(n+\frac{1}{3}\right)\lambda = R = aR'$, which leads to angles of local extrema as

$$\theta = \arccos(1 - \frac{(3n+1)^2}{18R'^2}) \tag{7}$$

Equations (6-7), applicable to hexagonal flake, are also summarized in Table 1. The explicit ordering of extrema can be easily tracked by considering the initial stacking.

The good agreements in the angles of local minimum, extracted from $S$ integral for hexagon of initial AA/AB stacking, and the scaling theories for hexagons (i.e., equations (6-7)), are apparent in Fig. 5(f-g). We also find that the angles of local minimum for the hexagon are slightly smaller (at most about 0.2 degree) than those for triangle at the same normalized flake size (see Section 6 in supporting information). This shows that flake shape may matter, although the deviation may be small.

The selection rules of twistronic angles for triangular and hexagonal 2D material flakes (equations (4-7)) are summarized in Table 1. For other flake shapes, the $S$ integral offers a general solution.

| Triangle | |
|---|---|
| AA $n^{th}$ local min | AB $n^{th}$ local min |
| $\theta = \arccos\left(1 - \frac{n^2}{2R'^2}\right)$ | $\theta = \arccos\left(1 - \frac{(2n+1)^2}{8R'^2}\right)$ |
| at $\theta = 1.1°$, $R' = 52.0879, 104.176, \ldots$ | at $\theta = 1.1°$, $R' = 73.1318, 130.22, \ldots$ |
| Hexagon | |
| AA $n^{th}$ local min | AB $n^{th}$ local min |
| $\theta = \arccos\left(1 - \frac{(2\sqrt{3}n-1)^2}{24R'^2}\right)$ | $\theta = \arccos\left(1 - \frac{(3n+1)^2}{18R'^2}\right)$ |
| at $\theta = 1.1°$, $R' = 37.0514, 89.1393, \ldots$ | at $\theta = 1.1°$, $R' = 69.4505, 121.538,\ldots$ |

Table 1. Selection rules of twistronic angles for triangular and hexagonal 2D material flakes as a function of dimensionless flake size for AA/AB stackings at the rotation center. A set of increasing magic-sizes (normalized) which prefers 1st magic-angle twist (1.1°) is calculated.

There are a few key implications for experimental efforts and observations based on the geometric selection rules. First, the intrinsic solution of accessing a particular desired twist angle may be addressed by fabricating the flake of required geometry using either chemical vapor deposition or etching, and then transferring the flake with a rough twist angle to a supporting layer. Then any perturbation (e.g. tip manipulation) would easily deliver a configuration with well-defined stable twist angles. Note that there are multiple orders of local minima that correspond to a single targeted twist angle, offering large designing space. Second, given the geometric conditions that derive these scaling laws, it is easy to see that higher-order angles of local minima are associated with more spatial periods of moiré pattern within the flake. In other words, the orders of local minima are linked to the topology of dislocation configuration (e.g., number of dislocation junctions) at the desired twist angle. Therefore, rotational stability [29] of these angles of local minima are protected by the topology of dislocation configuration. Third, these angles of local minima decrease and gradually approach zero as the flake size increases. Therefore, tiny twisted angles near or smaller than 1 degree are intrinsically accessible using flakes of larger sizes given the scaling laws, which can potentially offer a solution to engineering intrinsic twistronic structures targeting near-zero twist angles, such as the zero-approaching sequence of multiple magic-angles of graphene [8], given current challenge of accessing very tiny twist angles. In particular for the twisted bilayer graphene system, Table 1 presents a few magic sizes of the flake which intrinsically prefer the 1$^{st}$ magic-angle of 1.1 degree rotation as the local energy minimum. For other higher order magic-angle [8] (e.g, 0.5°, 0.35°, 0.24°, and 0.2°), the set of increasing magic-sizes can also be feasibly calculated. Note that for a given angle, there is always a set of increasing sizes that is available, which offers flexibility and feasibility for experimental investigation on a fast-increasing number of moiré periods within the flake (e.g., see Section 4 in supporting information).

**CONCLUSION**

In sum, we reveal a simple mapping between intrinsically preferred twist angles and geometry of the twisted bilayer system, in the form of geometric scaling laws for a wide range of intrinsically preferred twist angles as a function of only geometric parameters of the rotating flake on a supporting layer. These scaling laws are formulated for triangular or hexagonal shapes, as these are the most frequently encountered shapes in chemical vapor deposition growth. It is thus possible to envision an intrinsic twistronic structure, where the top layer flakes intrinsically favor the desired twist angles, the robustness and precision of which are protected by intrinsic interface energetics. Given the geometric nature of scaling laws, a general method for handling arbitrary flake shapes by integrating Fourier expansion of the periodic geometry is also proposed. Our analysis can be easily applied to any practical experimental systems. With this analysis, the energy-preferred twist angles during interlayer rotation in any 2D material bilayer systems shall become immediately clear. Our framework offers a general geometric solution to accessing a wide range of twist angle, including very tiny angles such as higher order zero-approaching sequence of multiple magic-angles of graphene [8] (e.g, 0.5°, 0.35°, 0.24°, and 0.2°). Given the strong connection of twist angles to various emerging condensed matter properties in 2D material layers, these scaling laws offer rich opportunities for the on-demand design and investigation of intrinsic "twistronics".

**ACKNOWLEDGEMENT**

We gratefully acknowledge the grants that supported this research. SZ acknowledges funding from Zhejiang University. EA and HJ also acknowledge the support of the Army Research Office (W911NF-17-1-0544) Material Science Division under Dr. Chakrapani Varanasi, and the partial support of NSF grant number CMMI 18-25300 (MOMS program).